\documentclass[english,nature,notitlepage,superscriptaddress,12pt]{revtex4-1}
\pdfoutput=1

\usepackage{color}

\usepackage{amsmath, amsthm, amsfonts}    
\usepackage{amssymb}

\usepackage[unicode=true,pdfusetitle,
 bookmarks=true,bookmarksnumbered=false,bookmarksopen=false,
 breaklinks=true,pdfborder={0 0 0},backref=false,colorlinks=true,citecolor=blue]{hyperref}

\usepackage{graphicx}   
\usepackage{subfig}       

\usepackage{bm}            
\usepackage[normalem]{ulem}  

\def\Rpart{\operatorname{Re}}
\def\Ipart{\operatorname{Im}}

\def\be{\begin{eqnarray}}
\def\ee{\end{eqnarray}}
\def\r{{\bf r}}

\def\E{{\bf E}}

\def\F{{\bf F}}

\def\im{{\rm i}}

\begin{document}

\title{Light induced ``Mock Gravity" at the nanoscale}

\author{J. Luis-Hita}
\email{jorge.luis@uam.es}
\affiliation{Donostia
International Physics Center (DIPC), Paseo Manuel Lardizabal 4, 20018
Donostia-San Sebastian, Spain}
\affiliation{
Departamento de F\'isica de la Materia Condensada,  Universidad Aut\'onoma de Madrid (UAM), 28049 Madrid, Spain}

\author{M.I. Marqu\'es}
\affiliation{
 Condensed Matter Physics Center (IFIMAC)  and Instituto ``Nicol\'as Cabrera'', Universidad Aut\'onoma de Madrid, 28049 Madrid, Spain}

\affiliation{
Departamento de F\'isica de Materiales,  UAM, 28049 Madrid, Spain}

\author{R. Delgado-Buscalioni}
\affiliation{
 Condensed Matter Physics Center (IFIMAC)  and Instituto ``Nicol\'as Cabrera'', Universidad Aut\'onoma de Madrid, 28049 Madrid, Spain}

\affiliation{
Departamento de F\'isica Te\'orica de la Materia Condensada, UAM, 28049 Madrid, Spain}

\author{N. de Sousa} 
\affiliation{Donostia
International Physics Center (DIPC), Paseo Manuel Lardizabal 4, 20018
Donostia-San Sebastian, Spain}

\author{L.S.\ Froufe-P\'erez}
\affiliation{Department of Physics, University of Fribourg, CH-1700 Fribourg, Switzerland}

\author{F. Scheffold}
\affiliation{Department of Physics, University of Fribourg, CH-1700 Fribourg, Switzerland}

\author{J.J.  S\'{a}enz}
\email{juanjo.saenz@dipc.org}
\affiliation{Donostia
International Physics Center (DIPC), Paseo Manuel Lardizabal 4, 20018
Donostia-San Sebastian, Spain}
\affiliation{IKERBASQUE, Basque Foundation for Science, 48013 Bilbao, Spain}

\date{\today}

\begin{abstract}
\end{abstract}
\maketitle
{\bf{
The origin  of  long-range attractive interactions has fascinated scientist along centuries.
The remarkable Fatio-LeSage's \cite{Fatio1690,lesage1784lucrece} corpuscular theory, introduced as early as in 1690 and generalized to electromagnetic waves by Lorentz \cite{lorentz1927lectures},  proposed that, due to their mutual shadowing,  two absorbing particles in an isotropic radiation field experience an attractive force which follows a gravity-like inverse square distance law. Similar ``Mock Gravity'' interactions were later introduced by Spitzer \cite{spitzer1941dynamics} and Gamow \cite{gamow1949relativistic} in the context of Galaxy formation but their actual relevance  in Cosmology has never been unambiguously established \cite{hogan1986galaxy,wang1989galaxy}.  Here we predict the existence of  Mock-Gravity,  ``$ \ 1/r^{2} \ $'', attractive forces between  two identical molecules or nanoparticles in a quasi monochromatic isotropic random light field,  whenever the light frequency is tuned to an absorption line such that the real part of the particle's electric polarizability is zero, i.e. at the  so-called  Fr\"ohlich resonance 
\cite{Bohren}. 
These interactions are scale independent, holding for both near and far-field separation distances.}}

The interaction between two objects  is usually defined to be long ranged if the force  decays with their distance apart, $r$, as a power law $ \sim 1/r^{n+1} \ $ with $n$ smaller than the spatial dimension of the system. Gravity is  a typical example of a long-range attractive force in three-dimensions while the interaction between electric or magnetic dipoles (n=3)   is borderline in between short and long range attraction  \cite{israelachvili2011intermolecular}.  In contrast, the familiar dispersion forces between non-polar, neutral, molecules and particles, arising from quantum electrodynamic fluctuations,  are short range. At close distances the Coulomb interaction between the fluctuating electric dipole moments leads to an interaction energy proportional to $1/r^6$, the so-called van deer Waals-London dispersion forces \cite{london1937general}. However,  when $r$ is larger than a characteristic resonance wavelength $\lambda_F$,  retardation effects become important since  the dipole moments fluctuate many times  over the period the light takes to pass between particles. The interaction energy varies then as $1/r^7$ as first shown by Casimir and Polder \cite{casimir1948influence}.  These interactions can also be derived as a special case of Lifshitz's theory of attraction between macroscopic bodies \cite{lifshitz1956theory}  in which the force is deduced from  {\em equilibrium} quantum and thermal electromagnetic field fluctuations \cite{lifshitz1956theory,mclachlan1963retarded,boyer1973retarded,henkel2002radiation}. 

In the last years there has been an increasing interest in understanding  the non-equilibrium analogs of Casimir  forces arising in the interaction between bodies at different temperature \cite{messina2011casimir,bimonte2017nonequilibrium} like those induced by  blackbody radiation from a hot source  on atoms and nanoparticles \cite{aunon2012optical,sonnleitner2013attractive}.
Surprisingly strong long-range interactions between atoms or non-absorbing dielectric particles  in a quasi-monochromatic fluctuating random field  were predicted \cite{Thiru80,SukhovDouglasDogariu2013,Brugger15} and experimentally demonstrated for micron-sized particles \cite{Brugger15}
(similar interactions   between pairs of dipoles under the excitation of multiple laser beams were also discussed \cite{Odell2000}). Although the effective interaction range can be controlled by the  spectral bandwidth of the fluctuating field \cite{Brugger15,rodriguez2009inter,holzmann2016tailored}  (with the Casimir-Lifshsitz interaction recovered  in the limit of a quantum black body spectrum \cite{Brugger15}), 
the existence of three-dimensional artificial gravity like, inverse square law, interaction forces had not yet been demonstrated. 

For non-absorbing dipolar particles in a quasi-monochromatic random field, the force always presents a characteristic oscillatory behaviour  for distances larger than the light wavelength (reminiscent of a Fabry-Perot-like behaviour).  Gravity-like interactions were predicted only for small separation distances  \cite{Thiru80,SukhovDouglasDogariu2013,Odell2000} assuming that the imaginary part of the polarizability could be  neglected, i.e. neglecting radiation pressure effects. However, as discussed below, radiation pressure effects dominate the near-field interactions of non-absorbing particles, leading to a rather different interaction law. Our main goal here is to show that, in contrast with atoms or dielectric particles, the interaction force between two identical  resonant molecules or plasmonic nanoparticles, whose extinction cross section is dominated by absorption,  can follow a true  attractive inverse square law all the way from near to far-field separation distances. As we will see, the ideal non-oscillating $ \sim 1/r^{2} \ $ law can only be achieved when the  frequency of the random field is tuned to the particles's Fr\"ohlich resonance (e.g. the Fr\"ohlich frequency, $\omega_F$, of plasmonic silver nanoparticles),  clarifying the  physical basis of the so-called Mock Gravity and  opening the possibility to study (mock) gravitational interactions at the nanoscale. Suspensions of  Fr\"ohlich  resonant nanoparticles will then offer a promising laboratory for testing the intriguing predictions of the statistical mechanics of systems with long-range interactions \cite{campa2009statistical}. 


To this end, let us consider two identical  nanospheres of radius $a$   separated by a distance $r$  in an otherwise  homogeneous  medium with  refractive index $n_h=1$. The particles  are illuminated by an homogeneous and isotropic random light field consisting  of a superposition of unpolarized and angularly uncorrelated plane waves (of frequency $\omega$ and wave number $k=\omega/c$ , being  $c$ is the vacuum speed of light). If the spheres are sufficiently small,  they can be characterized by 
their  electric polarizability $\alpha(\omega)$
\be
{\alpha}(\omega) &=& \left[
{\alpha}^{-1}_{0}(\omega) -\im \frac{k^3}{6\pi}\right]^{-1}
= 
|\alpha(\omega)| e^{i\delta_\omega}  \label{alpha1}  
\ee
where  ${\alpha}_{0}(\omega)$ is a quasistatic polarizability (real in absence of absorption) and $\delta_\omega$ the scattering phase-shift.
In order to discuss the ``$r$''-dependence of the   interaction force given by Eq. \eqref{Force5} in the  small particle limit, $ka \ll 1$,  we consider ${\alpha}_{0}(\omega)$  given by   
\be {\alpha}_{0}(\omega) = 4\pi a^3 \frac{\epsilon(\omega) -1}{\epsilon(\omega)+2} \label{alpha0}
\ee
where the particles' permittivity $\epsilon(\omega)$ is assumed  to follow  a Lorentz-Drude-like dispersion,
\be
\epsilon(\omega)=1+\frac{\omega_{p}^{2}}{\omega_{0}^{2}-\omega^{2}-\im\omega\Gamma_{0}}
\ee
(being $\omega_{p}$ the plasma frequency, $\omega_{0}$ the natural frequency and $\Gamma_{0}$ the damping constant).
The polarizability can then be written as \cite{carminati2006radiative,markel2007propagation}
\be
\alpha(\omega) &=& \frac{4\pi a^3 (\omega_F^2-\omega_{0}^{2})}{\omega_F^2-\omega^2 -\im \left\{\omega \Gamma_0 + 2 (ka)^3 (\omega_F^2-\omega_0^2)/3\right\} } \label{alfaLor} \ee
 where $\omega_F$ is the  Fr\"ohlich resonance frequency given by $\omega_F^{2}=\omega_p^{2}/3+\omega_0^2$. The term $-\im \Gamma_0\omega$ accounts for damping by absorption, whereas  $-\im 2 (ka)^3 (\omega_F^2-\omega_0^2)/3$ accounts for radiative damping \cite{carminati2006radiative}.

\begin{figure}
\centering

        \includegraphics[width=0.47\textwidth]{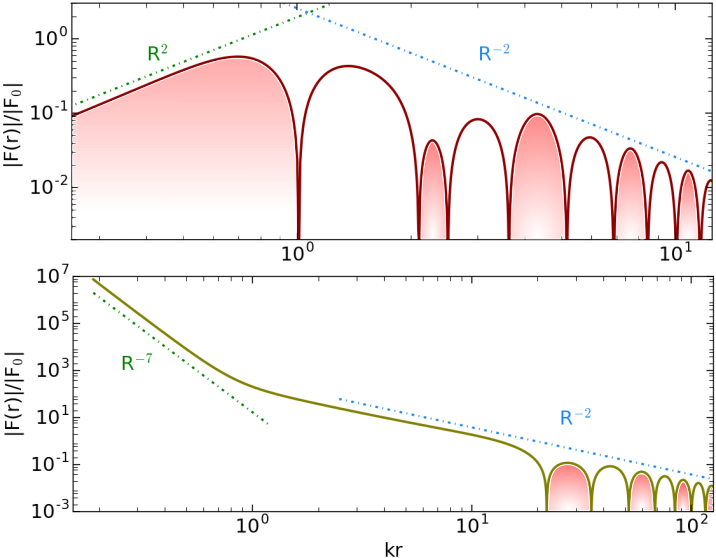}
    \hspace{0.02cm}
    
  \caption{{\bf Forces between non-absorbing particles.} Log-log plot of the absolute value of the interaction force for the Lorentz model with parameters $\omega_{0}=0.1\omega_F$, $\Gamma_{0}=0$ and $a=\lambda_F/100$ for (a) the resonant frequency $\omega=\omega_F$ and (b) strongly out of resonance $\omega=0.1\omega_F$. Red shadowed regions indicate repulsive interaction force.}
\label{Fig1}
\end{figure}

\begin{figure}
\centering

        \includegraphics[width=0.47\textwidth]{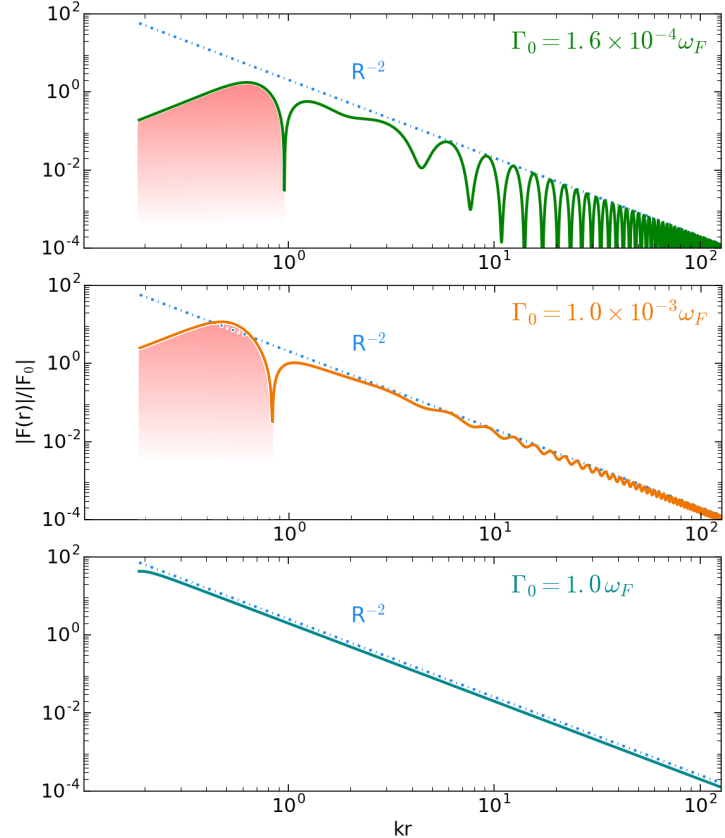}
    \hspace{0.02cm}
    
  \caption{log-log plot of the absolute value of the interaction force for the Lorentz model with parameters $\omega_{0}=0.1\omega_F$, $\omega=\omega_F$ and $a=\lambda_F/100$ for different damping coefficients (shown in each plot). Red shadowed regions indicate repulsive interaction force.}
\label{Fig2}
\end{figure}

In absence of absorption, $\sigma_{\text{abs}} =0$, the interaction force ( given by Eq. \eqref{Fnoabs} in Methods ),  exhibits an oscillatory behaviour in the far-field zone ($kr \gg 1$) with an envelop that decays 
as $r^{-2}$: 
\begin{align}
\left.  \lim_{kr \gg 1} \F_{12}(r)\right|_\text{No abs}
&\sim     - U_E \frac{k^4  | \alpha|^2}{2\pi} \frac{ \cos(2[kr +\delta_\omega])}{(kr)^2}
\frac{\r}{r},  \end{align}
a result that was first  predicted \cite{Thiru80} for the interactions between  dipolar particles  excited by a spatially coherent field after averaging over all orientations of the inter-atomic axis with respect to the incident beam (strictly equivalent to a fixed dimer illuminated by a random fluctuating field \cite{SukhovDouglasDogariu2013}). 
In the near-field zone, $kr \ll 1$,  we can distinguish two different regimes. 
At resonance, $\omega=\omega_F$, the  phase shift $\delta_\omega = \pi/2$ and $\alpha(\omega_F) = \im 6\pi k_F^{-3}$ and the near-field force is repulsive (see Methods), proportional to  the energy density of the random field, 
\begin{align}
\left. \lim_{ kr \ll 1} \F_{12}(r,\omega_F)\right|_\text{No abs} & \sim U_E \frac{k^4  | \alpha(\omega_F)|^2}{\pi} \frac{1}{3} (kr)^2 \ \frac{\r}{r}
\\ & = 12  \pi U_E  \  r^2 \frac{\r}{r} \end{align}
being independent  on the actual resonant frequency, $\omega_F$,  or any other particle's property. This universal limit had not been noticed previously.  In contrast,
in the weak scattering limit (strongly off-resonance) when $\omega \ll \omega_F$, as long as  
$a \ll r $, the interaction force goes as
\begin{align}
\left.  \lim_{ka \ll kr \ll 1} \F_{12}(r,\omega)\right|_\text{No abs} & \sim - U_E \frac{k^4  | \alpha|^2}{4\pi}  \times \nonumber \\ &\Bigg\{    \frac{22}{15}\frac{\cos(2\delta_\omega)}{(kr)^2}  
+ 18 \frac{\sin(2\delta_\omega) }{(kr)^7}  \Bigg\} \frac{\r}{r}
.
\end{align}
Previous works \cite{Thiru80,SukhovDouglasDogariu2013,Odell2000} disregard the last term assuming that, far from resonance, the imaginary part of the polarizability can be neglected (i.e.  $\sin 2 \delta_\omega \sim 2 \delta_\omega \sim 0$ and $\cos 2 \delta_\omega \sim 1$ ) which would lead to an attractive  $r^{-2}$, gravity-like, interaction force at short distances. However, even for frequencies  strongly  off-resonance ($\omega \ll \omega_F$) where $|\alpha| \sim 4\pi a^3$, in absence of absorption Optical Theorem imposes  $  \sin \delta_\omega = k^3 |\alpha|/(6\pi) \sim 2 (ka)^3/3 $, i.e. $\sin 2\delta_\omega \sim   2 \delta_\omega \sim 4 (ka)^3/3$.  This implies that, for small distances, the attractive term $\sim r^{-7}$ dominates the interaction. 
These results for non-absorbing particles are summarized in Fig.~\ref{Fig1} where we plot the force (normalized to $F_0 = U_E k^4  | \alpha(\omega)|^2/(4\pi)$, in our case $ F_0\simeq 10^{-18} $) versus separation distance for different illumination  frequencies.  Forces were calculated from Eq. \eqref{Force5} using the polarizability given by \eqref{alfaLor}. We compare the modulus of the actual force versus distance (in logarithmic scale) with the trends expected for $r^{2}$ and $r^{-2}$ in Figure 1.a (resonant case) and $r^{-7}$ and $r^{-2}$ in Figure 1.b (out of resonance). Note how a crossover from a $r^{-2}$  to a $r^{-7}$ tendency takes place as the particles get closer. Clearly, except for a narrow window of separation distances, in absence of absorption the interaction forces do not follow an attractive gravity like interaction.

\begin{figure}
\centering

        \includegraphics[width=0.4\textwidth]{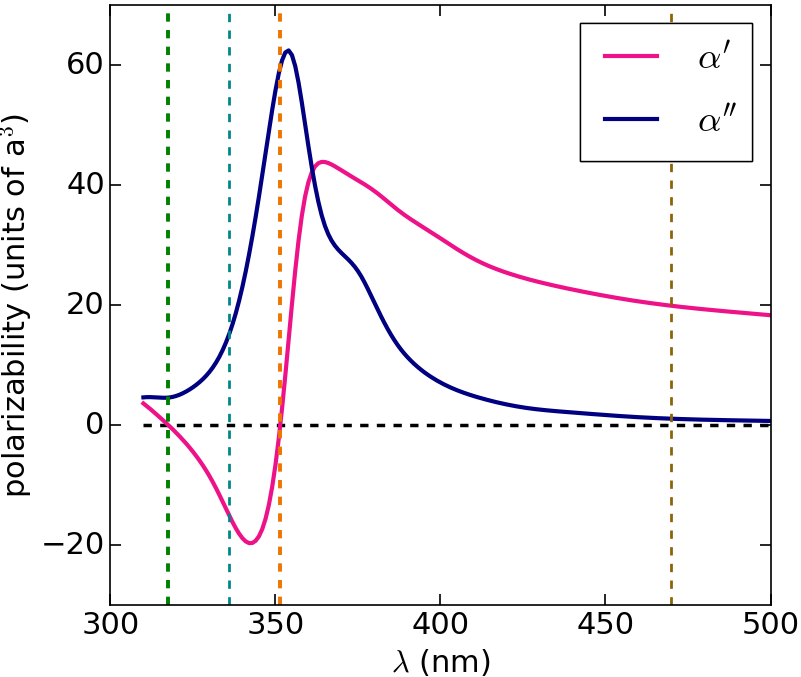}
    \hspace{0.02cm}
    
  \caption{Real  (black line) and imaginary part (red line) of the polarizability versus wavelength in vacuum for a silver nanoparticle with $a=5nm$. The interaction forces corresponding to  $\lambda=317, \ 337, \ 352$  and 470$nm$ (vertical dashed lines) are shown in Figure \ref{Fig4}. }
\label{Fig3}
\end{figure}

Let us now consider the forces for very small absorbing particles (e.g. few nm sized Ag particles \cite{coronado2011optical})  such that the extinction cross section is dominated by absorption \cite{Bohren},
i.e. $\sigma_{\text{abs}} \sim \sigma_{\text{ext}} $. In the weak scattering limit, the interaction force presents again an oscillatory behaviour  except at the Fr\"ohlich resonance, where the force can be shown to be given by \begin{align}
\left. 
  \F_{12}(r, \omega_F) \right|_{\text{abs}}
& \sim     - U_E \frac{k_F^4  | \alpha(\omega_F)|^2}{2\pi} 
\frac{1}{(k_Fr)^2} \ \frac{\r}{r}, \label{RESULT}
\end{align}
i.e. 
a force that is a non-oscillating  long range gravity-like interaction. This equation summarises  the most important result of the present work. 
Notice that within the small particle dipole approximation and for the Lorentz model, the weak scattering limit  at the  Fr\"ohlich resonance is given by
\be
 \left(\frac{a}{r}\right)^6  \left(\frac{\omega_F^{2}-\omega_{0}^{2}}{\Gamma_0\omega_F}\right)^2 \ll 1.
 \label{abscon} \ee 
and then Equation \eqref{RESULT} will hold 
 for distances as small as $r \sim 3 a$ (for shorter distances high order multipoles start being relevant) as long as the quality factor of the resonance $\frac{\omega_F^{2}-\omega_{0}^{2}}{\Gamma_0\omega_{F}}$ remains smaller than $\sim 30$. This is illustrated in Fig.~\ref{Fig2}, where we show the exact interaction force based on the polarizability given in Eq. \eqref{alfaLor} for different values of $\Gamma_0/\omega_F$. Under the condition given by Eq. \eqref{abscon}, the interaction turns gravitational-like at all separation distances,
all the way from the near to the far field zones.

\begin{figure}
\centering

        \includegraphics[width=0.47\textwidth]{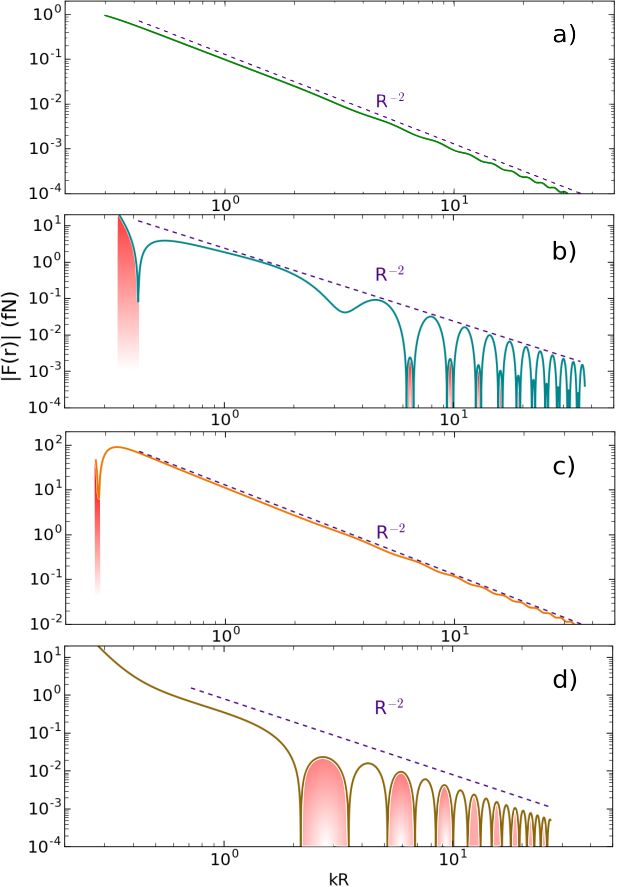}
    \hspace{0.02cm}
    
  \caption{{\bf Forces between Silver nanoparticles}. log-log plot of the interaction force for two silver nanoparticles with $a=5nm$ illuminated with a isotropic fluctuating random field of intensity $10W/\mu m^{2}$ for (a) $\lambda=317$nm, (b) $\lambda=$337nm, (c) $\lambda=352$nm and (d) $\lambda=470$nm. Red shadowed regions indicate repulsive interaction force.}
\label{Fig4}
\end{figure}

 In order to check the validity of the results in a realistic scenario, we consider the polarizability 
 given by Eqs. \eqref{alpha1} and \eqref{alpha0}  using experimental values for the permittivity of silver nanoparticles \cite{Palik}.
For simplicity,  we do not include corrections for nonlocal or size dependent dielectric response.
The polarizability of a $5$ nm radius silver nanoparticle in vacuum is represented in Fig.~\ref{Fig3}. Note how the real part of the polarizability is equal to zero at $317$ nm and $352$ nm.
Hence, gravity-like interactions should show up for the two wavelengths at which condition $\Rpart (\alpha)=0$ is fulfilled.
This is indeed what we observe in Fig.~\ref{Fig4} where we plot the interaction force given by Eq. \eqref{Force5} for the particular case of silver nanoparticles and compare with the expected behavior given by Eq. \eqref{RESULT}. 
Note how for $\lambda=317$ nm and $\lambda=352$nm the  gravitational-like interaction shows up, applying from infinity to short distances until absorption is not large enough to preserve the weak  scattering approximation. However, for a wavelength at which the  Fr\"ohlich condition is not fulfilled (for instance $\lambda=337$nm) the gravitational-like behavior disappears yielding to an oscillatory behaviour at long distances. It is worth to emphasise that, for example, gold nanoparticles would not present a clear  gravity-like interaction  since   the real part of the polarizability of a gold nanoparticle does not vanish.

\section*{Methods}

It can be shown that the averaged force on particle ``1'' located at $\r$, due to the presence of particle ``2''  at the origin of coordinates, can be written as the sum of two terms \cite{SukhovDouglasDogariu2013,Brugger15}:
\begin{subequations}
\begin{align}
 \F_{12}(r) 
&=     \Big\{\frac{ 4 \pi U_E}{k^2}\Big\} \sum_{i=x,y,z}  \Bigg[\Ipart\left\{  \frac{k^6 \alpha^2 g_i  g^\prime_i  }{ 1- k^6  \alpha^2 g^2_i}\right\} \label{Fnoabs}
  \\ & -     
 \Big\{  k^2 \sigma_{\text{abs}} \Big\} 
 \frac{\Rpart \left\{k^3 \alpha \left[g_i g_i^\prime + g_i {g_i^\prime}^*\right]\right\} }{ \left| 1- (k^3  \alpha)^2 g^2_i \right|^2} 
 \Bigg] \frac{\r}{r}
\end{align}
  \label{Force5}
\end{subequations}
where  $\sigma_{\text{abs}} = k \Ipart\{ \alpha\} - k^4 |\alpha|^2/(6\pi)$ is the absorption cross section of a single particle,  $U_E =   \epsilon_0  \left\langle |\E(\r,t)|^2 \right\rangle/2$  is the, time-averaged, energy of the fluctuating electric field per unit of volume ($U_E=U_{EM} /2$, being $U_{EM}$ the energy density of the electromagnetic wave),
and $g^\prime_i =\partial g_i/\partial (kr) $ with
\be
g_x(kr) &=& g_y(kr)= \frac{ e^{\im kr}}{4\pi kr} \left(1 + \frac{\im}{kr} - \frac{1}{(kr)^2} \right) \\
g_z(kr) &=& \frac{ e^{\im kr}}{4\pi kr} \left( -\frac{2 \im }{kr} + \frac{2}{(kr)^2} \right).
\ee
It is worth to mention that we implicitly assume that the system is in a stationary state and the energy absorbed is transferred to the thermal bath. Since the force is always directed along the radial direction and $\F_{12} = -\F_{21}$,  the force between two identical absorbing particles in a random, fluctuating field  is a {\em conservative force}. 

Equation   \eqref{Force5} simplifies considerably in the weak scattering limit , $|(k^3  \alpha) g_i|^2 \ll 1$, where  
recurrent scattering events do not play an relevant role (see the denominators in Eq. \eqref{Force5}).  
In absence of absorption, and strongly off-resonance ($\omega \ll \omega_F$), the weak scattering approximation holds even at near field distances ($kr \ll 1$),   
$|(k^3  \alpha) g_i|^2 
\sim (a/r)^6  \ll 1$, as long as  
$a \ll r $. However, at the resonance condition in absence of absorption $\alpha(\omega_F) = \im 6\pi k_F^{-3}$, and 
recurrent scattering  dominate the interaction force in the near field since 
 $\left|(k^3  \alpha) g_i\right|^2 \sim \left|6\pi g_i\right|^2 \gg 1 $. 

When the extinction cross section is dominated by absorption, $\sigma_{\text{abs}} \sim \sigma_{\text{ext}} $  the interaction force \eqref{Force5} in the  weak scattering limit is simply given by 
\begin{align}
\left.  
  \F_{12}(r) \right|_{\text{abs}}
& \sim      4\pi U_E  k^4  | \alpha|^2  \sum_{i=x,y,z} \Bigg[ \Ipart \left\{ \frac{ e^{2\im \delta_\omega} +1}{2}  g_i g_i^{\prime} \right\}
\nonumber \\ &  - \Ipart \left\{ \frac{ e^{2\im \delta_\omega} -1}{2}  g_i g_i^{\prime *} \right\} \Bigg]
 \ \frac{\r}{r}.  \end{align}
 At resonance $\omega=\omega_F$, $\delta_{\omega_F} = \pi/2$ and,
taking into account that
\be
 \sum_{i=x,y,z} \Ipart (g_i g_i^{\prime *})=\frac{-1}{8 \pi^2 (kr)^{2}},
 \ee
we obtain the interaction force given in Eq. \eqref{RESULT}.

\section*{Acknowledgments} This research was supported by the Spanish Ministerio de Econom\'ia y Competitividad (MICINN) and European Regional Development Fund (ERDF) through Projects  FIS2013-50510-EXP  and FIS2015-69295-C3-3-P, the Basque Dep.\ de Educaci\'on through Project PI-2016-1-0041,  the ``Mar\'ia de
Maeztu'' Program MDM-2014-0377 and the Swiss National Science Foundation through the National Center of Competence in Research Bio-Inspired Materials and through Project No.\ 149867 and 169074.


\section*{Author contributions} J.L.-H., M.I.M.  and J.J.S. conceived the study.  J.L.-H. carried out calculations and figures.   All authors contributed to the scientific discussion, writing and revising of the manuscript. J.J.S. supervised the study.


\section*{Competing financial interests} 
The authors declare no competing financial interests.


\bibliography{Light_Mock_Gravity.bib}

\begin{thebibliography}{10}
\expandafter\ifx\csname url\endcsname\relax
  \def\url#1{\texttt{#1}}\fi
\expandafter\ifx\csname urlprefix\endcsname\relax\def\urlprefix{URL }\fi
\providecommand{\bibinfo}[2]{#2}
\providecommand{\eprint}[2][]{\url{#2}}

\bibitem{Fatio1690}
\bibinfo{author}{Fatio~de Duillier, N.}
\newblock \bibinfo{title}{Correspondance with {Huygens}, {Letter} 2570}.
\newblock In \emph{\bibinfo{booktitle}{Oeuvres compl{\`e}tes de Christiaan
  Huygens. Correspondance, 1685-1690}}, \bibinfo{pages}{381--389}
  (\bibinfo{publisher}{M. Nijhoff (La Haye)}, \bibinfo{year}{1888}).

\bibitem{lesage1784lucrece}
\bibinfo{author}{LeSage, G.~L.}
\newblock \bibinfo{title}{Lucr{\`e}ce {Newtonien}}.
\newblock In \emph{\bibinfo{booktitle}{Nouveaux m{\'e}moires de l'Acad{\'e}mie
  royale des sciences et belles-lettres}}, \bibinfo{pages}{404--432}
  (\bibinfo{publisher}{George Jacques Decker, Berlin}, \bibinfo{year}{1784}).

\bibitem{lorentz1927lectures}
\bibinfo{author}{Lorentz, H.~A.}
\newblock \emph{\bibinfo{title}{Lectures on Theoretical Physics}}
  (\bibinfo{publisher}{Macmillan and Co., Limited. London},
  \bibinfo{year}{1927}).

\bibitem{spitzer1941dynamics}
\bibinfo{author}{Spitzer~Jr, L.}
\newblock \bibinfo{title}{The dynamics of the interstellar medium. ii.
  radiation pressure.}
\newblock \emph{\bibinfo{journal}{The Astrophysical Journal}}
  \textbf{\bibinfo{volume}{94}}, \bibinfo{pages}{232} (\bibinfo{year}{1941}).

\bibitem{gamow1949relativistic}
\bibinfo{author}{Gamow, G.}
\newblock \bibinfo{title}{On relativistic cosmogony}.
\newblock \emph{\bibinfo{journal}{Reviews of Modern Physics}}
  \textbf{\bibinfo{volume}{21}}, \bibinfo{pages}{367} (\bibinfo{year}{1949}).

\bibitem{hogan1986galaxy}
\bibinfo{author}{Hogan, C.} \& \bibinfo{author}{White, S.}
\newblock \bibinfo{title}{Galaxy formation by mock gravity}.
\newblock \emph{\bibinfo{journal}{Nature}} \textbf{\bibinfo{volume}{321}},
  \bibinfo{pages}{575--578} (\bibinfo{year}{1986}).

\bibitem{wang1989galaxy}
\bibinfo{author}{Wang, B.} \& \bibinfo{author}{Field, G.~B.}
\newblock \bibinfo{title}{Galaxy formation by mock gravity with dust ?}
\newblock \emph{\bibinfo{journal}{The Astrophysical Journal}}
  \textbf{\bibinfo{volume}{346}}, \bibinfo{pages}{3--11}
  (\bibinfo{year}{1989}).

\bibitem{Bohren}
\bibinfo{author}{{B}ohren, C.~F.} \& \bibinfo{author}{{H}uffman, D.~R.}
\newblock \emph{\bibinfo{title}{{A}bsorption and {S}cattering of {L}ight by
  {S}mall {P}articles}} (\bibinfo{publisher}{{W}iley-VCH Verlag GmbH},
  \bibinfo{address}{{B}erlin}, \bibinfo{year}{2007}).

\bibitem{israelachvili2011intermolecular}
\bibinfo{author}{Israelachvili, J.~N.}
\newblock \emph{\bibinfo{title}{Intermolecular and surface forces}}
  (\bibinfo{publisher}{Academic Press, Oxford}, \bibinfo{year}{2011}).

\bibitem{london1937general}
\bibinfo{author}{London, F.}
\newblock \bibinfo{title}{The general theory of molecular forces}.
\newblock \emph{\bibinfo{journal}{Transactions of the Faraday Society}}
  \textbf{\bibinfo{volume}{33}}, \bibinfo{pages}{8b--26}
  (\bibinfo{year}{1937}).

\bibitem{casimir1948influence}
\bibinfo{author}{Casimir, H.} \& \bibinfo{author}{Polder, D.}
\newblock \bibinfo{title}{The influence of retardation on the london-van der
  waals forces}.
\newblock \emph{\bibinfo{journal}{Physical Review}}
  \textbf{\bibinfo{volume}{73}}, \bibinfo{pages}{360} (\bibinfo{year}{1948}).

\bibitem{lifshitz1956theory}
\bibinfo{author}{Lifshitz, E.}
\newblock \bibinfo{title}{The theory of molecular attractive forces between
  solids}.
\newblock \emph{\bibinfo{journal}{Soviet Physics JETP}}
  \textbf{\bibinfo{volume}{2}}, \bibinfo{pages}{73--83} (\bibinfo{year}{1956}).

\bibitem{mclachlan1963retarded}
\bibinfo{author}{McLachlan, A.}
\newblock \bibinfo{title}{Retarded dispersion forces between molecules}.
\newblock In \emph{\bibinfo{booktitle}{Proceedings of the Royal Society of
  London A: Mathematical, Physical and Engineering Sciences}}, vol.
  \bibinfo{volume}{271}, \bibinfo{pages}{387--401} (\bibinfo{organization}{The
  Royal Society}, \bibinfo{year}{1963}).

\bibitem{boyer1973retarded}
\bibinfo{author}{Boyer, T.~H.}
\newblock \bibinfo{title}{Retarded van der waals forces at all distances
  derived from classical electrodynamics with classical electromagnetic
  zero-point radiation}.
\newblock \emph{\bibinfo{journal}{Physical Review A}}
  \textbf{\bibinfo{volume}{7}}, \bibinfo{pages}{1832} (\bibinfo{year}{1973}).

\bibitem{henkel2002radiation}
\bibinfo{author}{Henkel, C.}, \bibinfo{author}{Joulain, K.},
  \bibinfo{author}{Mulet, J.-P.} \& \bibinfo{author}{Greffet, J.-J.}
\newblock \bibinfo{title}{Radiation forces on small particles in thermal near
  fields}.
\newblock \emph{\bibinfo{journal}{Journal of Optics A: Pure and Applied
  Optics}} \textbf{\bibinfo{volume}{4}}, \bibinfo{pages}{S109}
  (\bibinfo{year}{2002}).

\bibitem{messina2011casimir}
\bibinfo{author}{Messina, R.} \& \bibinfo{author}{Antezza, M.}
\newblock \bibinfo{title}{Casimir-lifshitz force out of thermal equilibrium and
  heat transfer between arbitrary bodies}.
\newblock \emph{\bibinfo{journal}{EPL (Europhysics Letters)}}
  \textbf{\bibinfo{volume}{95}}, \bibinfo{pages}{61002} (\bibinfo{year}{2011}).

\bibitem{bimonte2017nonequilibrium}
\bibinfo{author}{Bimonte, G.}, \bibinfo{author}{Emig, T.},
  \bibinfo{author}{Kardar, M.} \& \bibinfo{author}{Kr{\"u}ger, M.}
\newblock \bibinfo{title}{Nonequilibrium fluctuational quantum electrodynamics:
  Heat radiation, heat transfer, and force}.
\newblock \emph{\bibinfo{journal}{Annual Review of Condensed Matter Physics}}
  \textbf{\bibinfo{volume}{8}}, \bibinfo{pages}{119--143}
  (\bibinfo{year}{2017}).

\bibitem{aunon2012optical}
\bibinfo{author}{Au{\~n}{\'o}n, J.~M.} \& \bibinfo{author}{Nieto-Vesperinas,
  M.}
\newblock \bibinfo{title}{Optical forces on small particles from partially
  coherent light}.
\newblock \emph{\bibinfo{journal}{JOSA A}} \textbf{\bibinfo{volume}{29}},
  \bibinfo{pages}{1389--1398} (\bibinfo{year}{2012}).

\bibitem{sonnleitner2013attractive}
\bibinfo{author}{Sonnleitner, M.}, \bibinfo{author}{Ritsch-Marte, M.} \&
  \bibinfo{author}{Ritsch, H.}
\newblock \bibinfo{title}{Attractive optical forces from blackbody radiation}.
\newblock \emph{\bibinfo{journal}{Physical Review Letters}}
  \textbf{\bibinfo{volume}{111}}, \bibinfo{pages}{023601}
  (\bibinfo{year}{2013}).

\bibitem{Thiru80}
\bibinfo{author}{Thirunamachandran, T.}
\newblock \bibinfo{title}{Intermolecular interactions in the presence of an
  intense radiation field}.
\newblock \emph{\bibinfo{journal}{Molecular Physics}}
  \textbf{\bibinfo{volume}{40}}, \bibinfo{pages}{393--399}
  (\bibinfo{year}{1980}).

\bibitem{SukhovDouglasDogariu2013}
\bibinfo{author}{{S}ukhov, S.}, \bibinfo{author}{{D}ouglas, K.~M.} \&
  \bibinfo{author}{{D}ogariu, A.}
\newblock \bibinfo{title}{{D}ipole-dipole interaction in random electromagnetic
  fields}.
\newblock \emph{\bibinfo{journal}{{O}ptics {L}etters}}
  \textbf{\bibinfo{volume}{38}}, \bibinfo{pages}{2385--2387}
  (\bibinfo{year}{2013}).

\bibitem{Brugger15}
\bibinfo{author}{{B}r\"ugger, G.}, \bibinfo{author}{{F}roufe {P}\'erez, L.~S.},
  \bibinfo{author}{{S}cheffold, F.} \& \bibinfo{author}{{S}\'aenz, J.~J.}
\newblock \bibinfo{title}{{C}ontrolling dispersion forces between small
  particles with artificially created random light fields}.
\newblock \emph{\bibinfo{journal}{{N}ature {C}ommunications}}
  \textbf{\bibinfo{volume}{6}}, \bibinfo{pages}{7460} (\bibinfo{year}{2015}).

\bibitem{Odell2000}
\bibinfo{author}{O'dell, D.}, \bibinfo{author}{Giovanazzi, S.},
  \bibinfo{author}{Kurizki, G.} \& \bibinfo{author}{Akulin, V.}
\newblock \bibinfo{title}{Bose-einstein condensates with 1/r interatomic
  attraction: Electromagnetically induced “gravity”}.
\newblock \emph{\bibinfo{journal}{Physical Review Letters}}
  \textbf{\bibinfo{volume}{84}}, \bibinfo{pages}{5687} (\bibinfo{year}{2000}).

\bibitem{rodriguez2009inter}
\bibinfo{author}{Rodr{\'\i}guez, J.} \& \bibinfo{author}{Andrews, D.~L.}
\newblock \bibinfo{title}{Inter-particle interaction induced by broadband
  radiation}.
\newblock \emph{\bibinfo{journal}{Optics Communications}}
  \textbf{\bibinfo{volume}{282}}, \bibinfo{pages}{2267--2269}
  (\bibinfo{year}{2009}).

\bibitem{holzmann2016tailored}
\bibinfo{author}{Holzmann, D.} \& \bibinfo{author}{Ritsch, H.}
\newblock \bibinfo{title}{Tailored long range forces on polarizable particles
  by collective scattering of broadband radiation}.
\newblock \emph{\bibinfo{journal}{New Journal of Physics}}
  \textbf{\bibinfo{volume}{18}}, \bibinfo{pages}{103041}
  (\bibinfo{year}{2016}).

\bibitem{campa2009statistical}
\bibinfo{author}{Campa, A.}, \bibinfo{author}{Dauxois, T.} \&
  \bibinfo{author}{Ruffo, S.}
\newblock \bibinfo{title}{Statistical mechanics and dynamics of solvable models
  with long-range interactions}.
\newblock \emph{\bibinfo{journal}{Physics Reports}}
  \textbf{\bibinfo{volume}{480}}, \bibinfo{pages}{57--159}
  (\bibinfo{year}{2009}).

\bibitem{carminati2006radiative}
\bibinfo{author}{Carminati, R.}, \bibinfo{author}{Greffet, J.-J.},
  \bibinfo{author}{Henkel, C.} \& \bibinfo{author}{Vigoureux, J.-M.}
\newblock \bibinfo{title}{Radiative and non-radiative decay of a single
  molecule close to a metallic nanoparticle}.
\newblock \emph{\bibinfo{journal}{Optics Communications}}
  \textbf{\bibinfo{volume}{261}}, \bibinfo{pages}{368--375}
  (\bibinfo{year}{2006}).

\bibitem{markel2007propagation}
\bibinfo{author}{Markel, V.~A.} \& \bibinfo{author}{Sarychev, A.~K.}
\newblock \bibinfo{title}{Propagation of surface plasmons in ordered and
  disordered chains of metal nanospheres}.
\newblock \emph{\bibinfo{journal}{Physical Review B}}
  \textbf{\bibinfo{volume}{75}}, \bibinfo{pages}{085426}
  (\bibinfo{year}{2007}).

\bibitem{coronado2011optical}
\bibinfo{author}{Coronado, E.~A.}, \bibinfo{author}{Encina, E.~R.} \&
  \bibinfo{author}{Stefani, F.~D.}
\newblock \bibinfo{title}{Optical properties of metallic nanoparticles:
  manipulating light, heat and forces at the nanoscale}.
\newblock \emph{\bibinfo{journal}{Nanoscale}} \textbf{\bibinfo{volume}{3}},
  \bibinfo{pages}{4042--4059} (\bibinfo{year}{2011}).

\bibitem{Palik}
\bibinfo{author}{Palik, E.}
\newblock \emph{\bibinfo{title}{1985 Handbook of Optical Constants of Solids}}
  (\bibinfo{publisher}{Academic Press, New York}, \bibinfo{year}{1985}).

\end{thebibliography}

\end{document}